\documentclass{PoS}
\pdfoutput=1
\usepackage{wrapfig}
\usepackage{amsbsy,amssymb,latexsym,amsfonts,amsmath}
\usepackage{graphicx,color}

\title{Toward the Global Structure of Conformal Theories in the $SU(3)$ Gauge Theory}

\ShortTitle{Toward the Global Structure of Conformal Theories}

\author{
   Y. Iwasaki
      \thanks{presented by K. -I. Ishikawa on behalf of Y. Iwasaki:
      based on the work in collaboration with K. -I Ishikawa, Yu Nakayama and T. Yoshie}
   \\
   \\
   \\
   \llap{}
Center for Computational Sciences, University of Tsukuba, Tsukuba, Ibaraki 305-8577, Japan\\
      E-mail:\email{iwasaki@ccs.tsukuba.ac.jp}
}

\abstract{
We introduce a new concept ``conformal theories with an IR cutoff',
after pointing out that the following two categories in $SU(3)$ gauge theories with fundamental $N_f$ fermions  possess an IR fixed point: Large $N_f  (N_f^{c} \le N_f \le 16)$ QCD within the conformal window  (referred as Conformal QCD)  and 
small $N_f  (1 \le N_f \le N_f^{c}-1)$ QCD at temperature $T/T_c > 1$  (referred as High Temperature QCD).
In the case of Conformal QCD in the continuum limit, the compact space and/or time gives an IR cutoff.
In the case of High Temperature QCD, the temperature $T$ plays a role of an IR cutoff together with a cutoff due to possible compact space, depending on how to take the continuum limit.
We note any lattice calculation  is performed on a finite lattice. Thus any calculation on a lattice possesses an IR cutoff.

In the conformal theories with an IR cutoff there exists the ``conformal region'' together with the confining region and the deconfining region. We verify numerically on a lattice of the size $16^3\times 64$ the existence of the conformal region and the non-trivial $Z(3)$ structure of the vacuum and the Yukawa-type decay form of meson propagators in the conformal region.

We stress that High Temperature QCD is intrinsically accompanied with an IR cutoff. Therefore the understanding the vacuum structure and the property of correlation functions is the key to resolve long standing issues in High Temperature QCD.

We further argue that there is a precise correspondence between
 Conformal QCD and High Temperature QCD in the temporal propagators under the
change of the parameters $N_f$ and $T/T_c$ respectively: the one boundary is
close to meson states and the other is close to free quark states.

In particular, we find the correspondence between Conformal QCD with
$N_f = 7$ and High Temperature QCD with $N_f=2$ at $T\sim 2\, T_c$ being in
close relation to a meson unparticle model.
From this we estimate the anomalous mass dimension $\gamma^* = 1.2 (1)$ for $N_f=7$.
}

\FullConference{31st International Symposium on Lattice Field Theory - LATTICE 2013\\
		July 29 - August 3, 2013\\
		Mainz, Germany}

\begin{document}

\section{Strategy, Objectives and Summary}
Recently much attention has been paid to four dimensional conformal theories, since conformal theories or nearly conformal theories are attractive candidates for the beyond standard model.
To confront the nature, it is important to understand each conformal theory, and for this purpose,
it is urgent to clarify the global structure of conformal theories~\cite{review1}.

 In this article, we investigate the global structure of $SU(3)$ conformal theories with $N_f$ fundamental fermions,
on a lattice using the Wilson fermion. Since we have reported the results in detail~\cite{coll-full}, I will try to make
this writeup is complementary to ~\cite{coll-full}, giving a r\'{e}sum\'{e} without detailed discussion.

We first point out that
the following two categories in $SU(3)$ gauge theories with fundamental $N_f$ fermions  possess an IR fixed point:
\begin{itemize}
\item
Large $N_f  (N_f^{c} \le N_f \le 16)$ QCD within the conformal window  (referred as Conformal QCD); $N_f^c$ is the lower critical flavor number for the conformal window.
\item
small $N_f  (2 \le N_f \le N_f^{c}-1)$ QCD at temperature $T/T_c > 1$ with $T_c$ being the critical temperature (referred as High Temperature QCD) 
\end{itemize}

We then introduce a new concept ``conformal theories with an IR cutoff''~\cite{coll1}.
In the case of Conformal QCD in the continuum limit, the compact space and/or time gives an IR cutoff.
In the case of High Temperature QCD, the temperature $T$ plays a role of an IR cutoff together with a cutoff due to possible compact space, depending on how to take the continuum limit.
We note any lattice calculation  is performed on a finite lattice. Thus any calculation on a lattice possesses an IR cutoff.

Finally the objectives of this article are (for details of numerical simulations see Ref.~\cite{coll-full}.)
\begin{enumerate}
\item
Verify numerically on a lattice of the size $16^3\times 64$ that
the ``conformal region'' exists together with the confining region and the deconfining region
in the phase structure parametrized by $\beta$ and $K$,
both in Conformal QCD and  High Temperature QCD.

Further verify the vacuum of the conformal region is 
the nontrivial $Z(3)$ twisted vacuum modified by non-perturbative effects
and temporal propagators of meson behave at large $t$ as a power-law corrected Yukawa-type decaying form.
The transition from the conformal region to the deconfining region or the confining region
is  a transition between different vacua and therefore the transition is a first order transition
both in Conformal QCD and in High Temperature QCD.

\item
Verify a precise correspondence between Conformal QCD and High Temperature QCD within the conformal region is realized under the change of a continuous parameter $T/T_c$  and a discrete parameter $N_f$, respectively: the one boundary is close to meson states and the other is close to free quark states.
\end{enumerate}

We stress that High Temperature QCD is intrinsically accompanied with an IR cutoff. Therefore the understanding the vacuum structure and the property of correlation functions is the key to resolve long standing issues in High Temperature QCD.

\begin{figure*}[thb]
 \includegraphics [width=7.5cm]{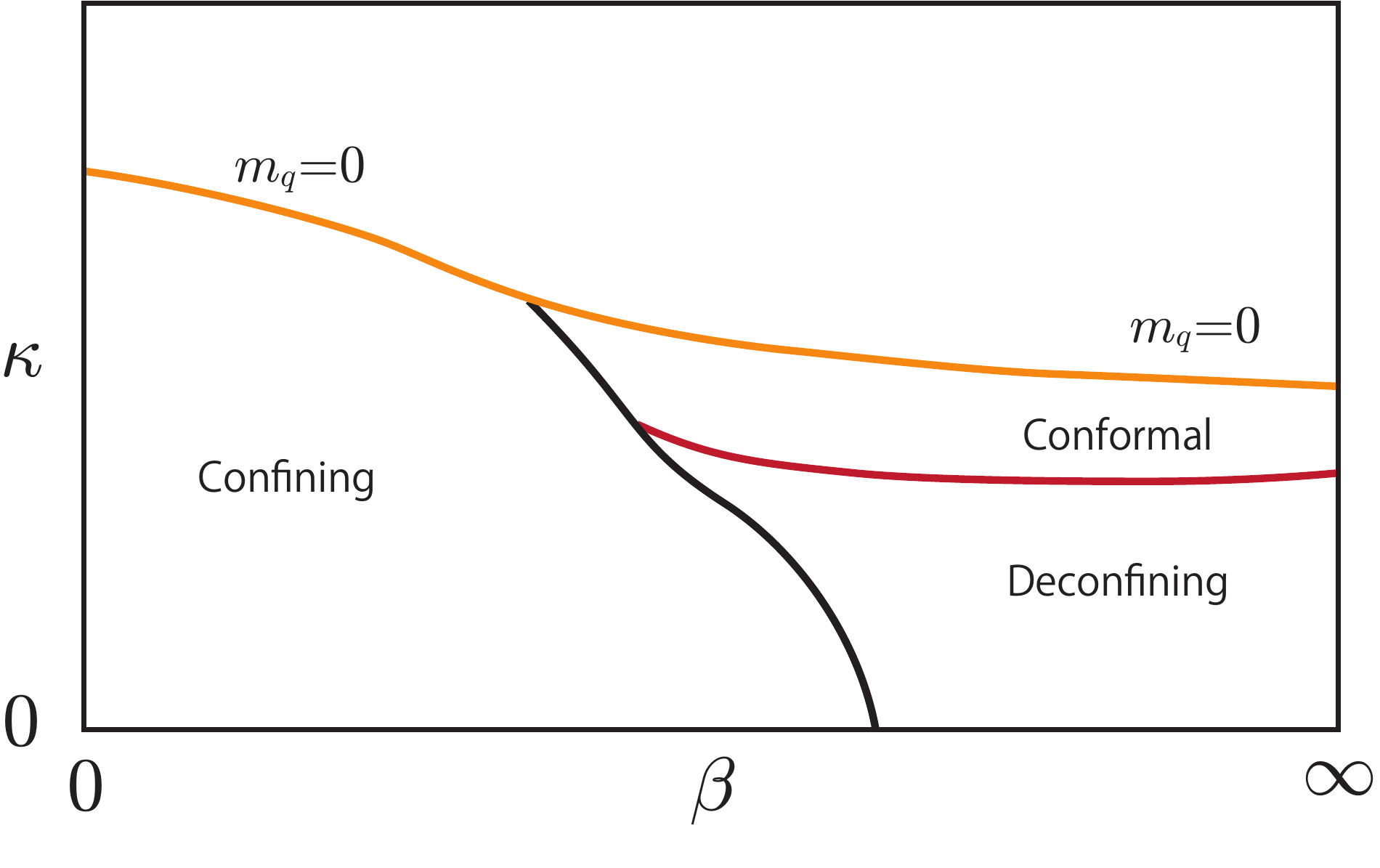}
 \hspace{1cm}
 \includegraphics [width=7.5cm]{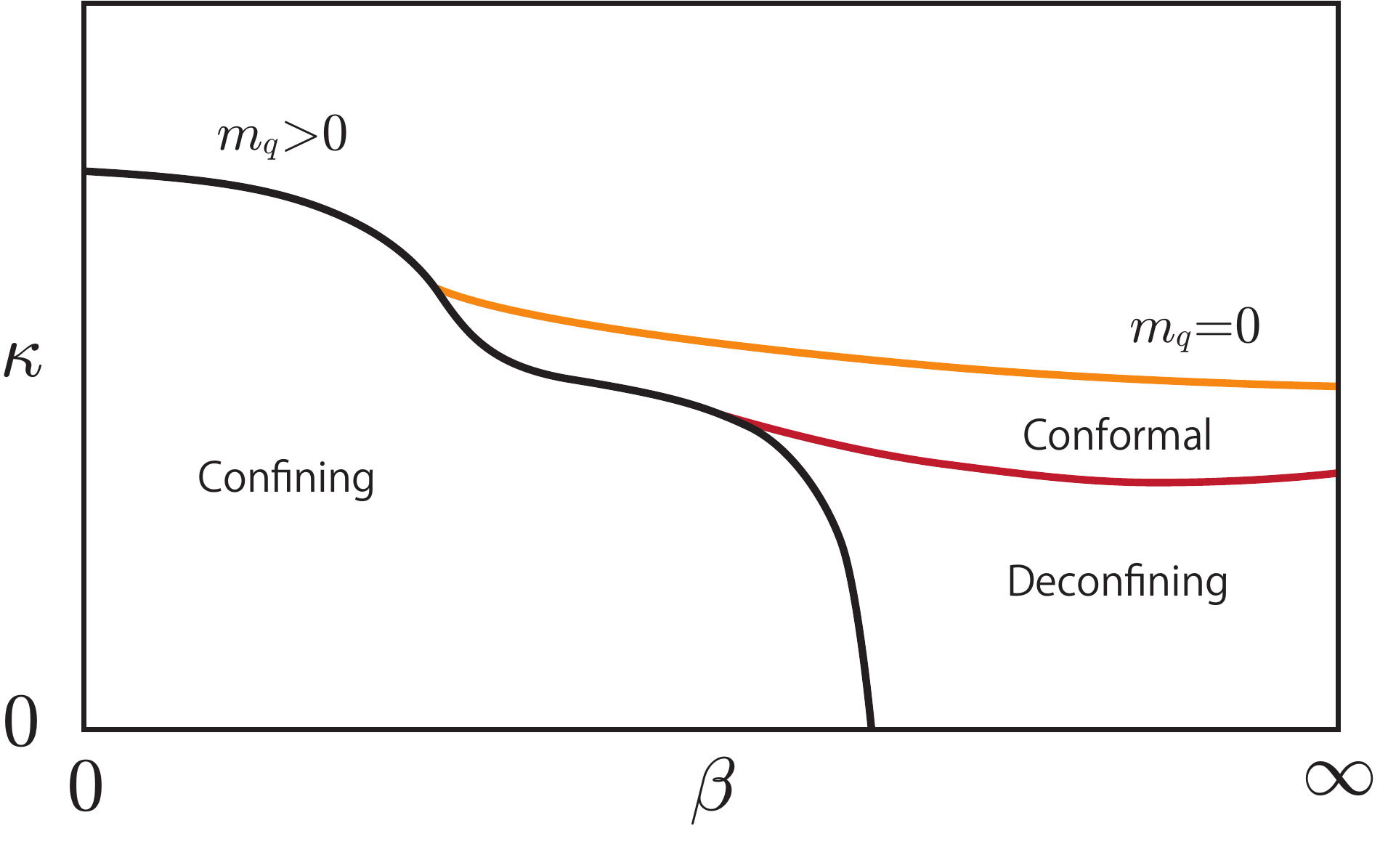}
\caption{ Phase diagram on a finite lattice : (left) $1 \le N_f \le N_f^c -1$ ; (right) $N_f^c \le N_f \le 16$;
 In the case $N_f^c \le N_f \le 16$ the massless quark line originating from the UV fixed point hits the bulk transition point at finite $\beta$ and no massless line exists in the confining region.  The region above the bulk transition corresponds to the doublers of the Wilson fermion. On the other hand, in the case  $1 \le N_f \le N_f^c -1$
 on the massless quark line there is a chiral phase transition point. Below the critical point the massless line is in the confining region.}
\label{phase diagram finite lattice}
\end{figure*}

\section{The existence of an IR fixed point}
The existence of an IR fixed point in Conformal QCD is well known as the Banks-Zaks IR fixed point 
\cite{Banks1982}.

In High Temperature QCD the existence of an IR fixed point has been recently pointed out  in Ref.~\cite{coll2}. 
Define a running coupling constant $g(\mu; T)$  at temperature $T$ in the massless quark case~(See for example \cite{karsch}).
When $T/T_c  > 1$, where the quark is not confined, the running coupling constant $g(\mu; T)$ cannot be arbitrarily large. This means that there is an IR fixed point with non-trivial zero of the beta  function when $T/T_c  > 1$.
This is the key observation. For example,
numerical results of the running coupling constant $g(r; T)$ shown in Fig. 2 in \cite{karsch} are consistent with
the above: the running coupling constant $g(r; T)$ increases as $r$ increases up to some value and does not further increases more than that.

The existence of an IR fixed point is not common in the lattice community.
One possible reason might be due to the non-vanishing trace anomaly in High Temperature QCD.
To make clear the implication of vanishing of the beta function at finite temperature, we recall the relation between the trace anomaly of energy momentum tensor and the beta function with massless quarks: 
$$
\langle T^{\mu}_{\ \mu} \rangle|_T = \mathcal{B}(g^{-2}(\mu))  \langle \mathrm{Tr}(F_{\mu\nu} (\mu))^2 \rangle|_T \ , 
$$
where $\mathcal{B}(g^{-2}(\mu))$ is the zero temperature beta function evaluated at $g = g(\mu)$, and 
$\langle \mathrm{Tr}(F_{\mu\nu} (\mu))^2 \rangle|_T$ is the field strength squared at temperature $T$ renormalized at scale $\mu$
(Appendix B of~\cite{coll-full}).
 
\begin{figure*}[t]
\includegraphics [width=15cm]{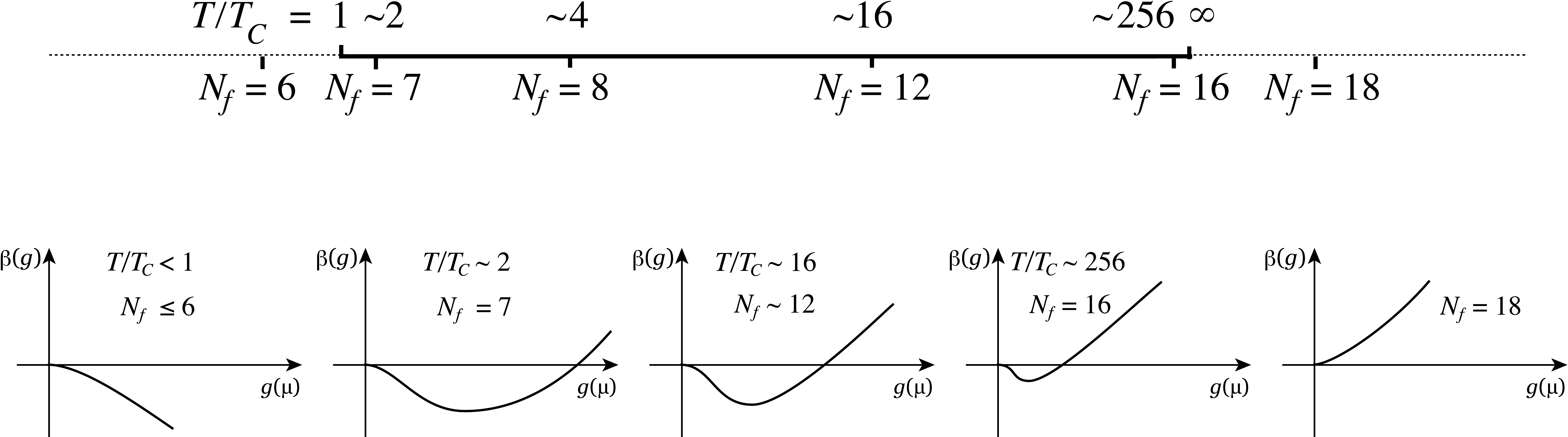}
\caption{Correspondence between Conformal QCD and High Temperature QCD in terms of the beta function.
The horizontal line on the top represents the correspondence between the number of flavor $N_f$ and the temperature $T/T_c$. }
\label{betaf2}
\end{figure*}

In Lorentz invariant zero-temperature field theories, the vanishing
beta function means that the trace anomaly vanishes and the theory is conformal invariant.
 However, when the beta
function at finite temperatures vanishes,
it does not  imply vanishing of the trace  of the energy-momentum tensor.
 Thus the vanishing beta function at $T>T_c$ does not contradict with the non-vanishing of
the difference of energy density and three times the pressure.

Another reason for that the existence of an IR fixed point is not common  might be due to the existence of an intrinsic IR cutoff in High Temperature QCD. We will discuss such cases below.

\section{Conformal theories with an IR cutoff and Conformal Region}
We first define the conformal field theories with an IR cutoff. The first assumption is that the beta function (either zero-temperature or finite temperature) vanishes. Of course, if there were no dimensionful quantities, this would imply that the theory is scale invariant and all the correlation functions show a strict power behavior.

Our new observation is that when such theories have a finite cutoff, then they will show the universal behavior that we call ``conformal field theories with an IR cutoff":
the ``conformal region'' exists together with the confining region and the deconfining region
in the phase structure.

We have verified numerically on a lattice of the size $16^3\times 64$ the existence of the conformal region
for $N_f=7, 8 , 12$ and $16$ as depicted on the right panel of Fig.~\ref{phase diagram finite lattice}
and for $N_f=2$ as on the left panel.

In the conformal region we find the vacuum is 
the nontrivial $Z(3)$ twisted vacuum modified by non-perturbative effects
and temporal propagators of meson behave at large $t$ as a power-law corrected Yukawa-type decaying form.
The transition from the conformal region to the deconfining region or the confining region
is  a transition between different vacua and therefore the transition is a first order transition
both in Conformal QCD and in High Temperature QCD.

\begin{figure*}[htb]
\includegraphics [width=8cm]{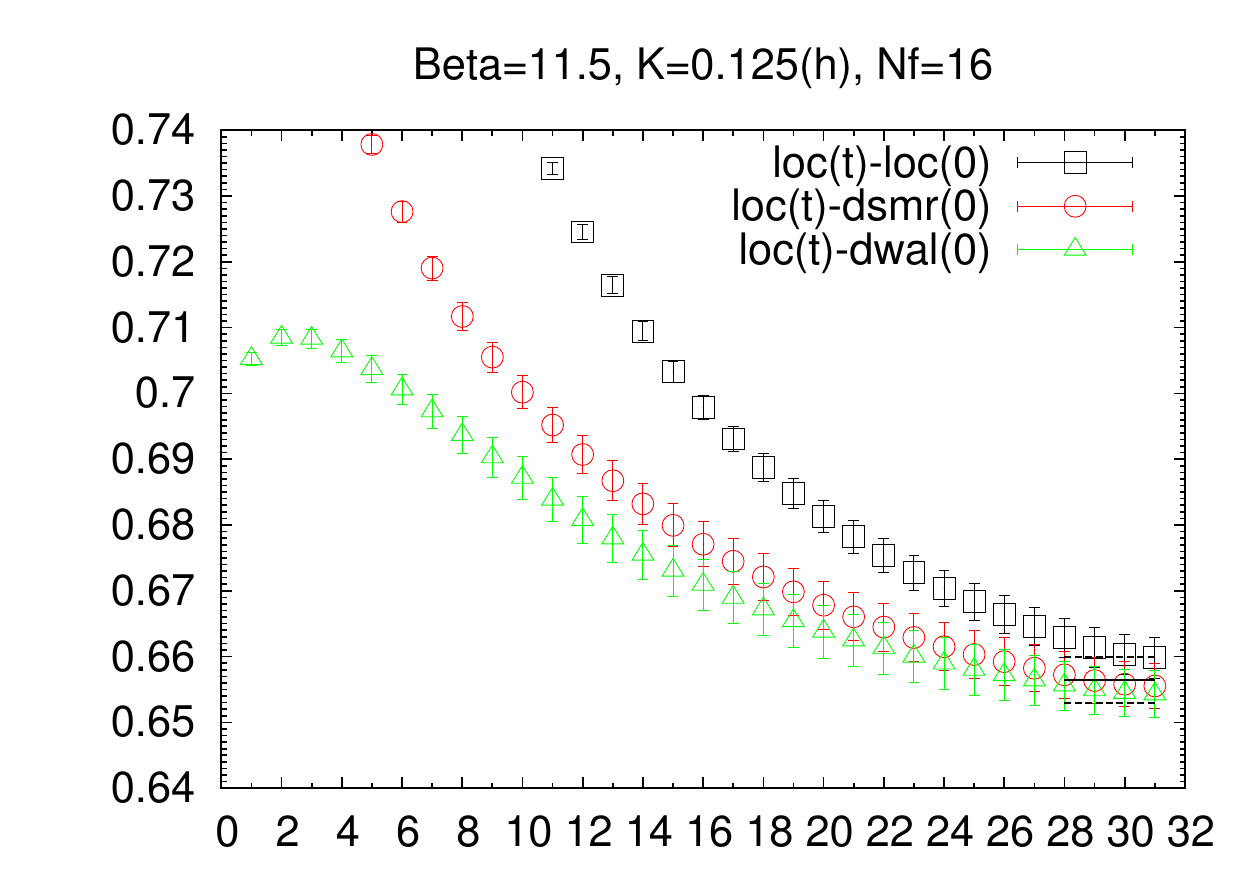}
\includegraphics [width=8cm]{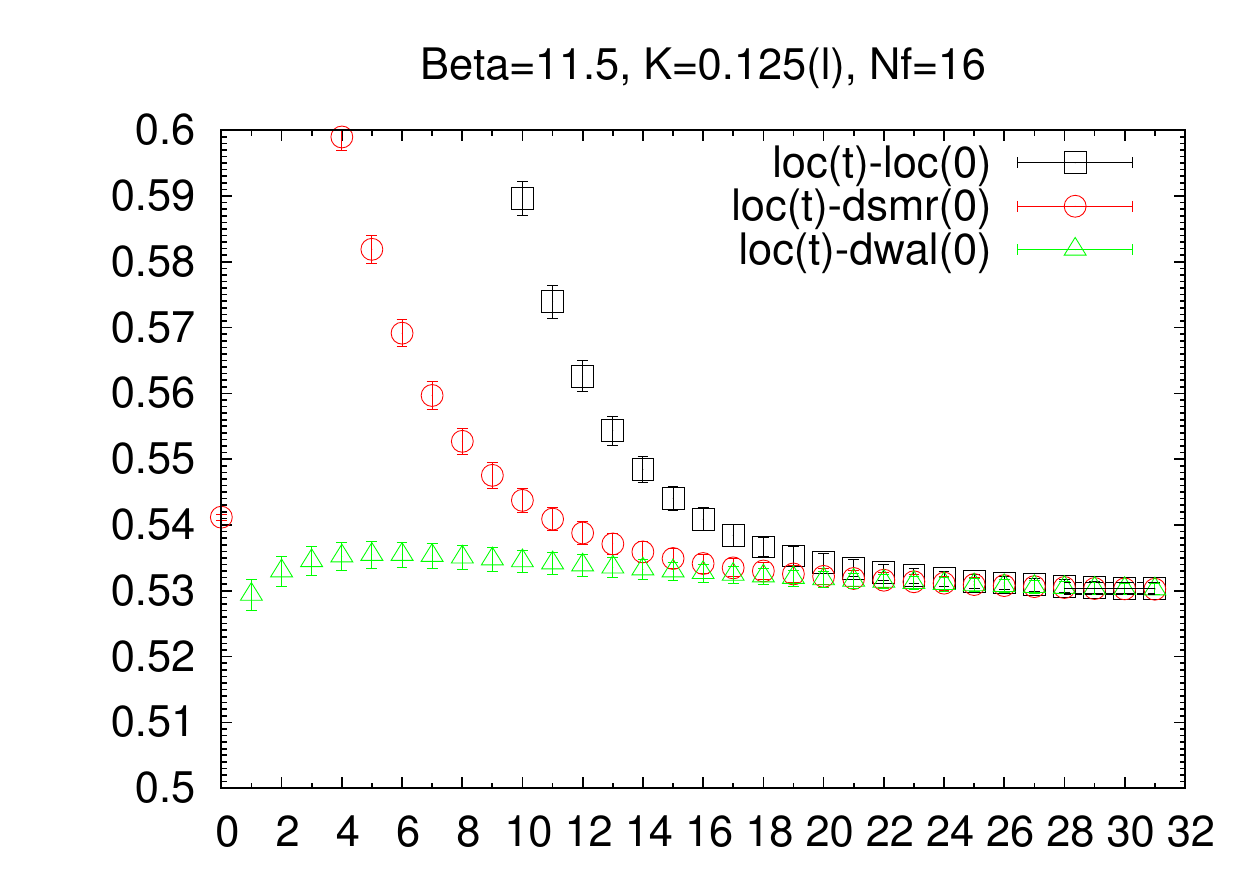}
\caption{ The effective mass: both for $N_f=16$ at $\beta=11.5$ and $K=0.125$: (left) from larger $K$ and (right) from smaller $K$;
See the main text for the three types of sources.}
\label{nf16_effm}
\end{figure*}

\begin{figure*}[htb]
\includegraphics [width=7.5cm]{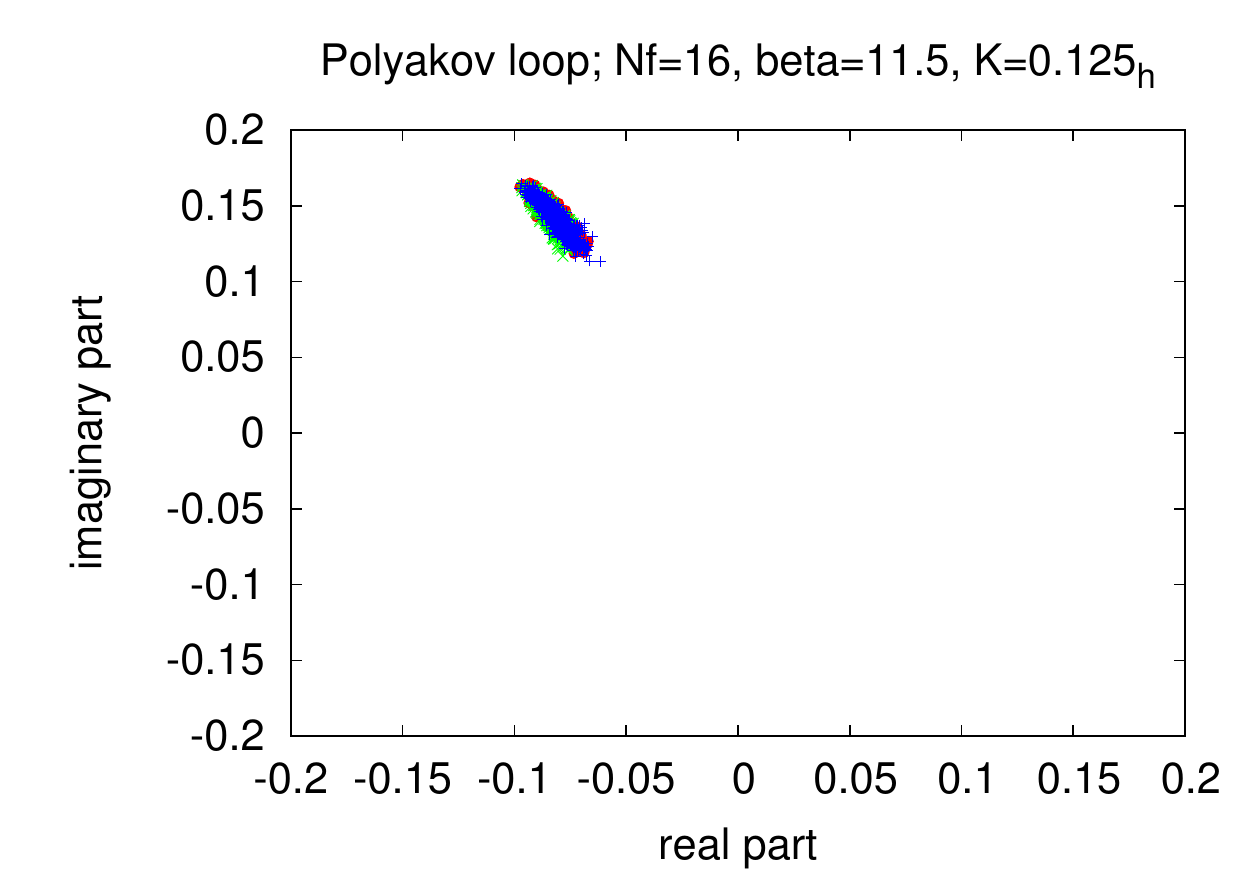}
 \hspace{1cm}
\includegraphics [width=7.5cm]{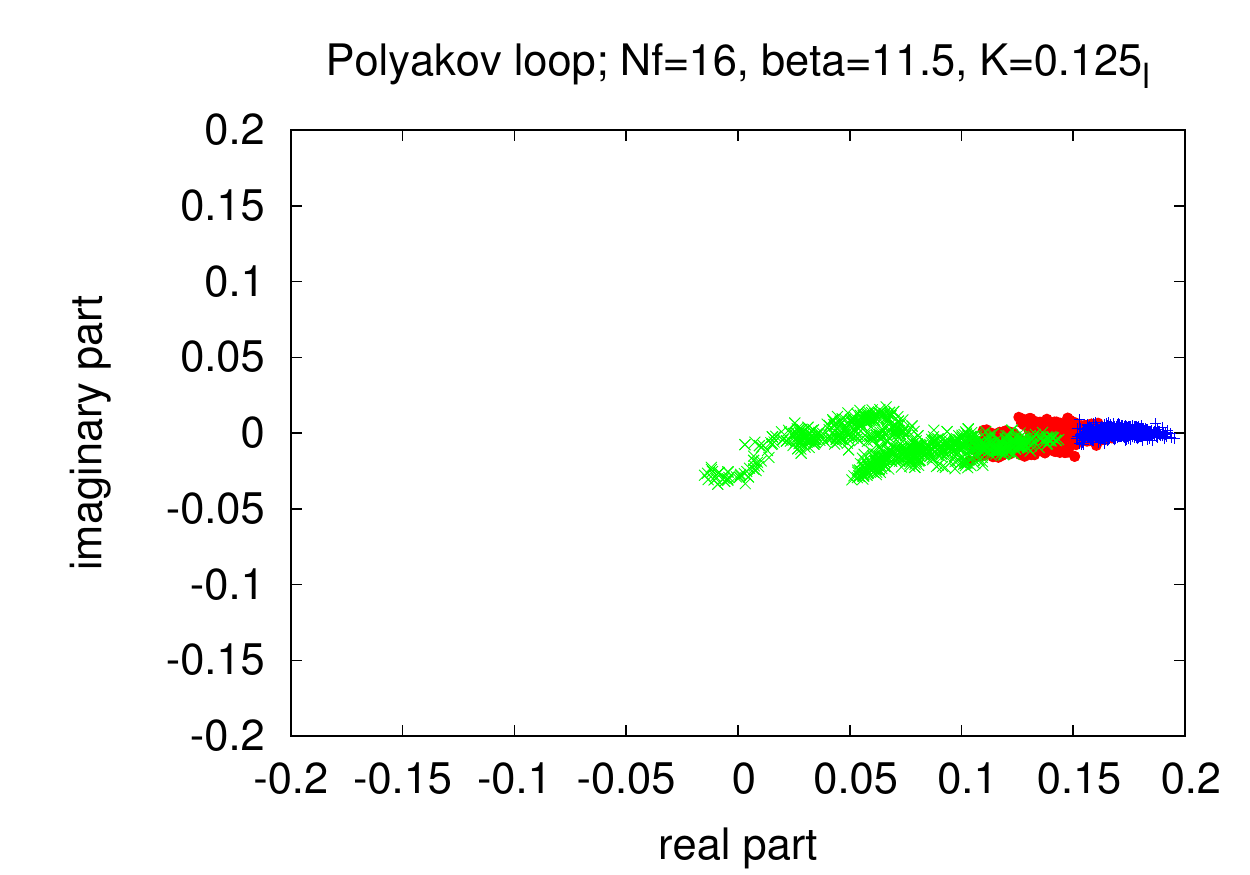}
\caption{ The scattered plots of Polyakov loops in the $x$, $y$ and $z$ directions overlaid; 
        both for $N_f=16$ at $\beta=11.5$ and $K=0.125$: (left) from larger $K$ and (right) from smaller $K$.}
\label{nf16_comp}
\end{figure*}

\begin{figure*}[thb]
\includegraphics [width=6.7cm]{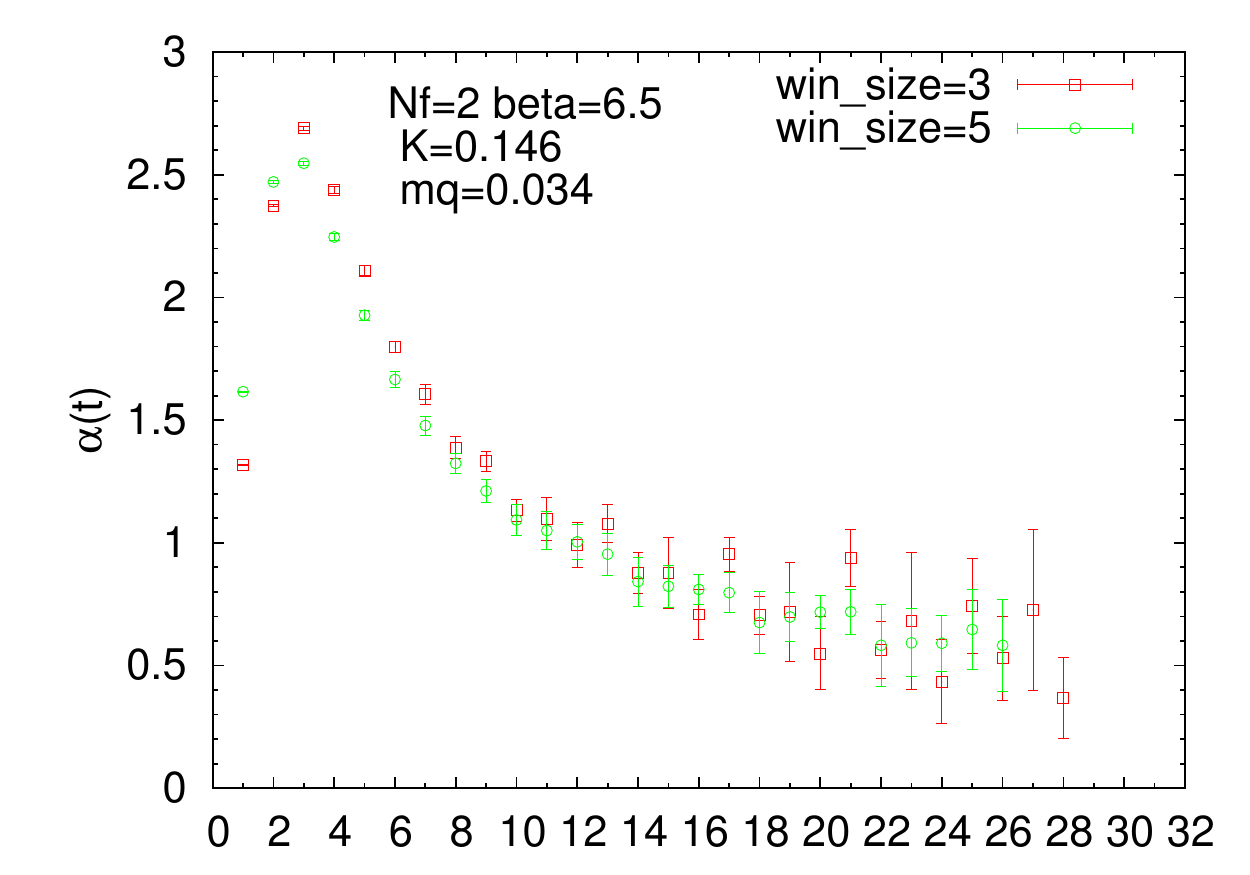}
\hspace{1cm}
\includegraphics [width=6.7cm]{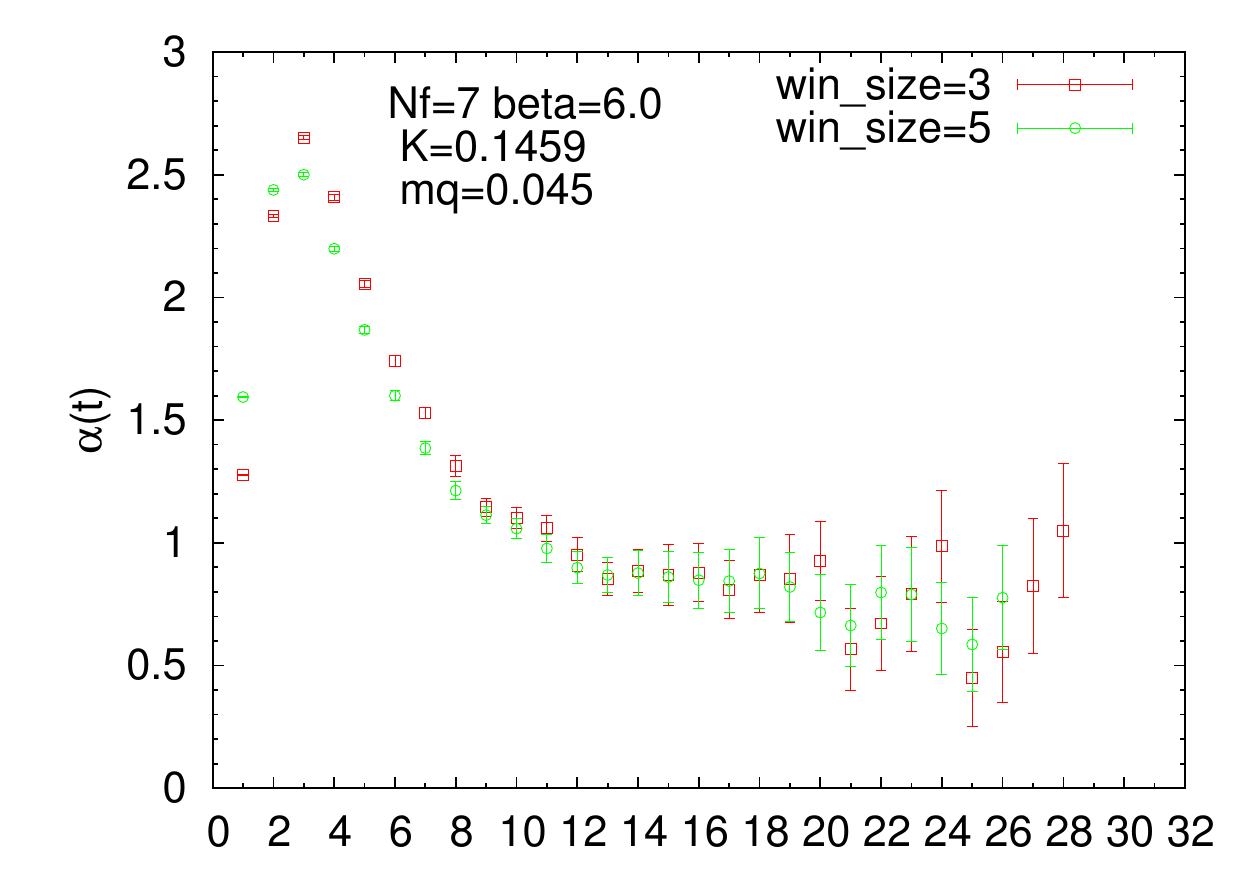}
\includegraphics [width=6.7cm]{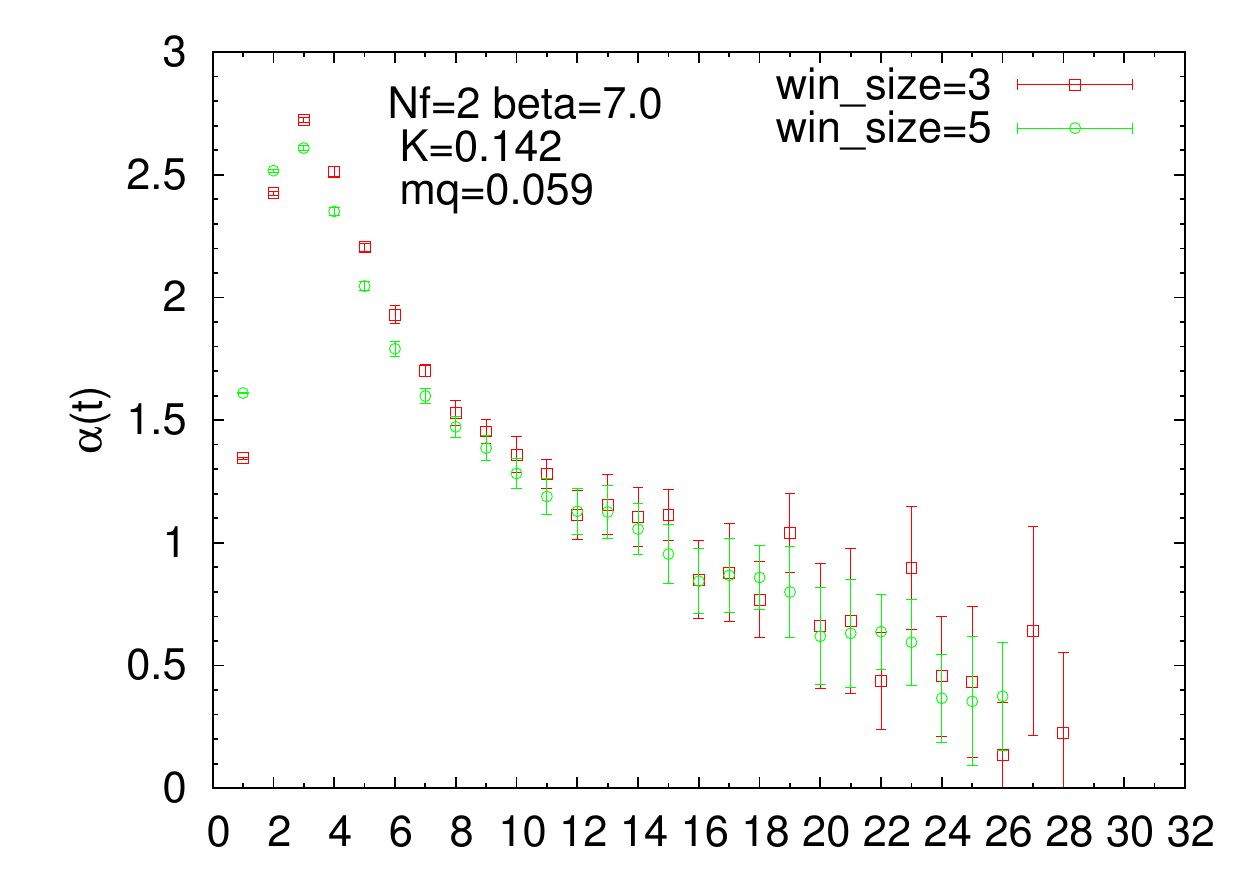}
\hspace{1cm}
\includegraphics [width=6.7cm]{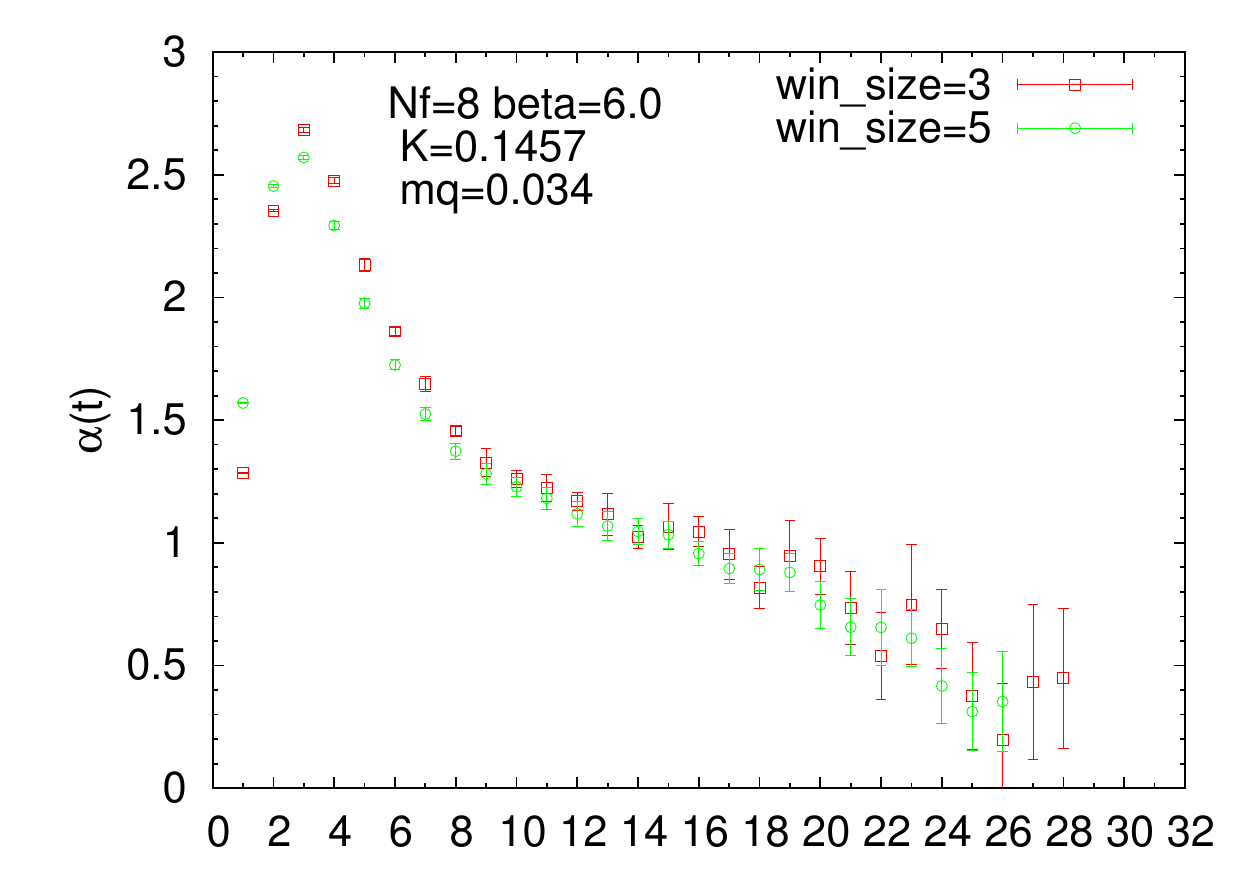}
\includegraphics [width=6.7cm]{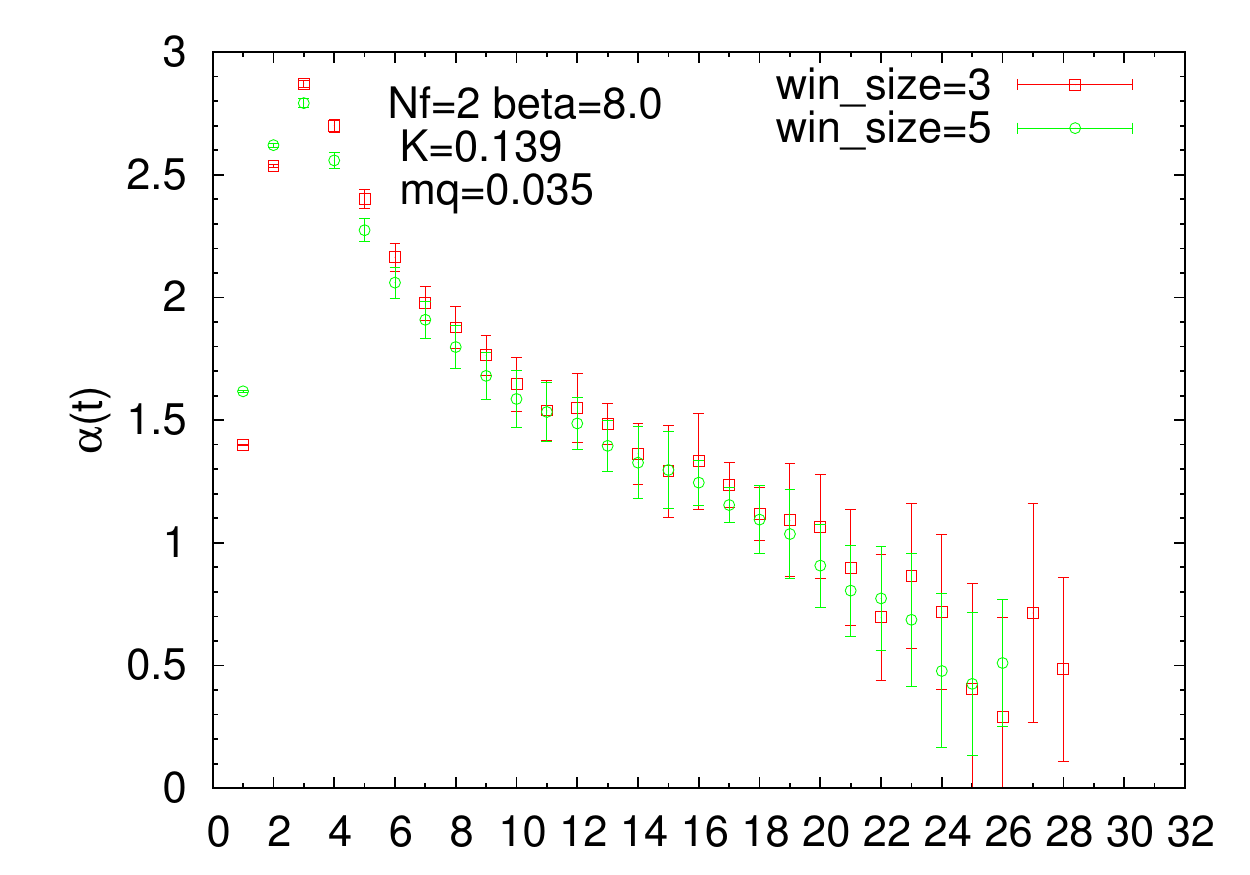}
\hspace{1cm}
\includegraphics [width=6.7cm]{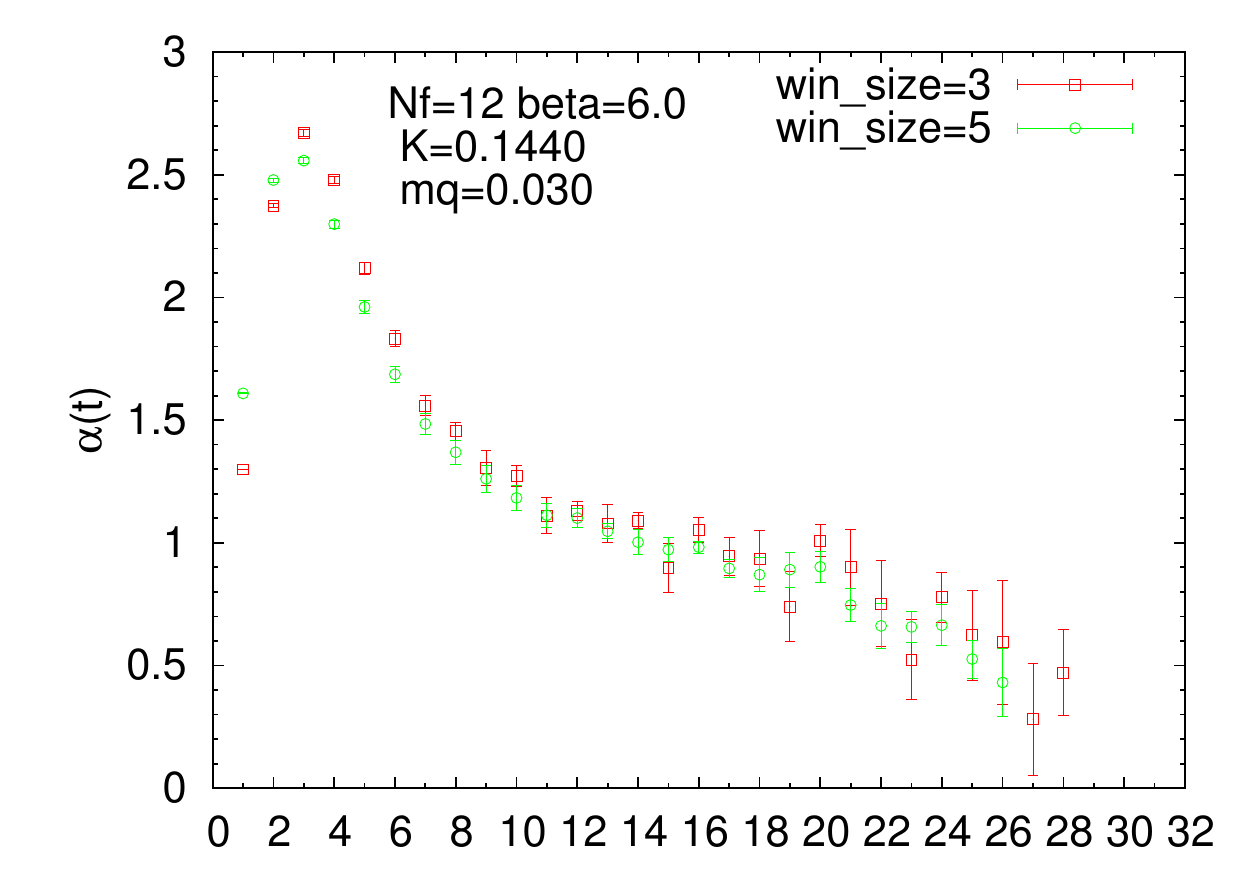}
\includegraphics [width=6.7cm]{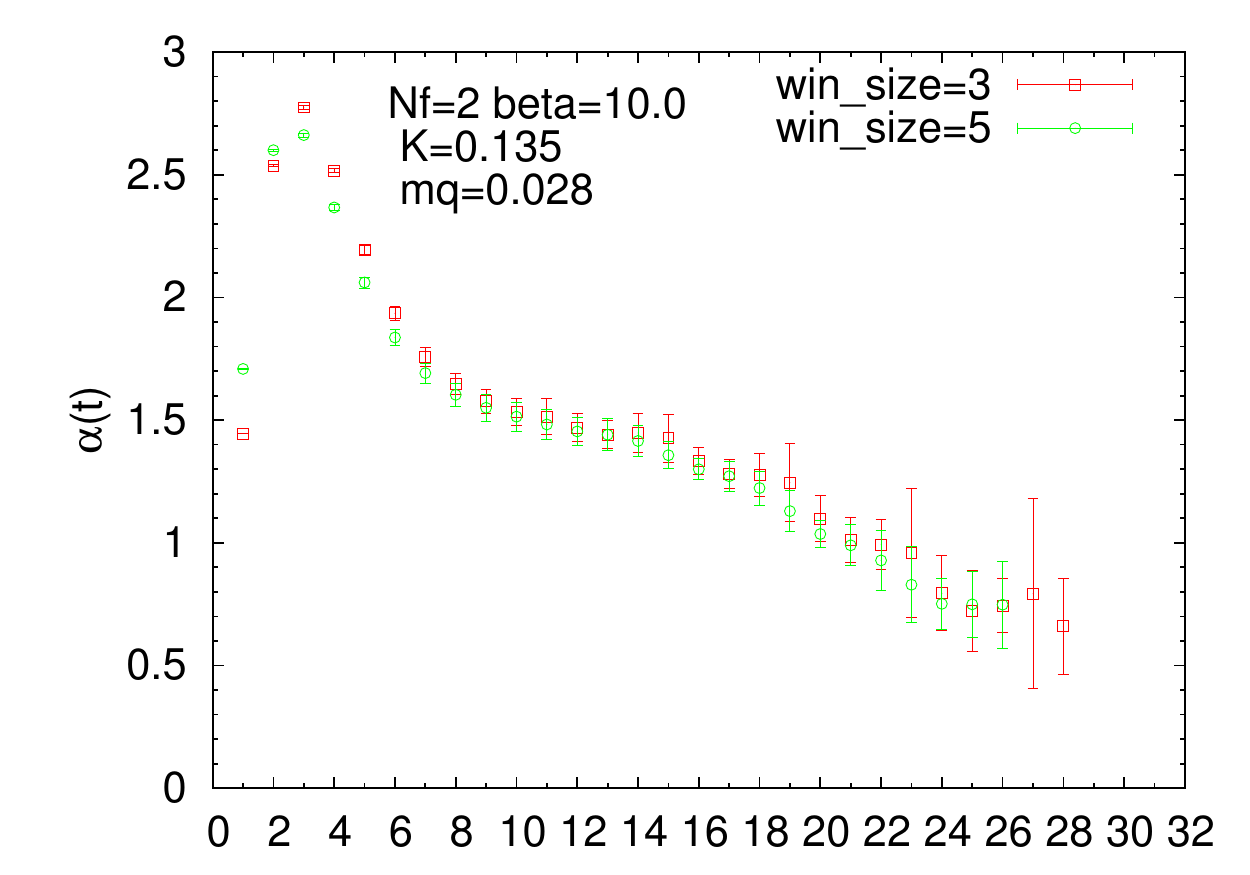}
\hspace{1.5cm}
\includegraphics [width=6.7cm]{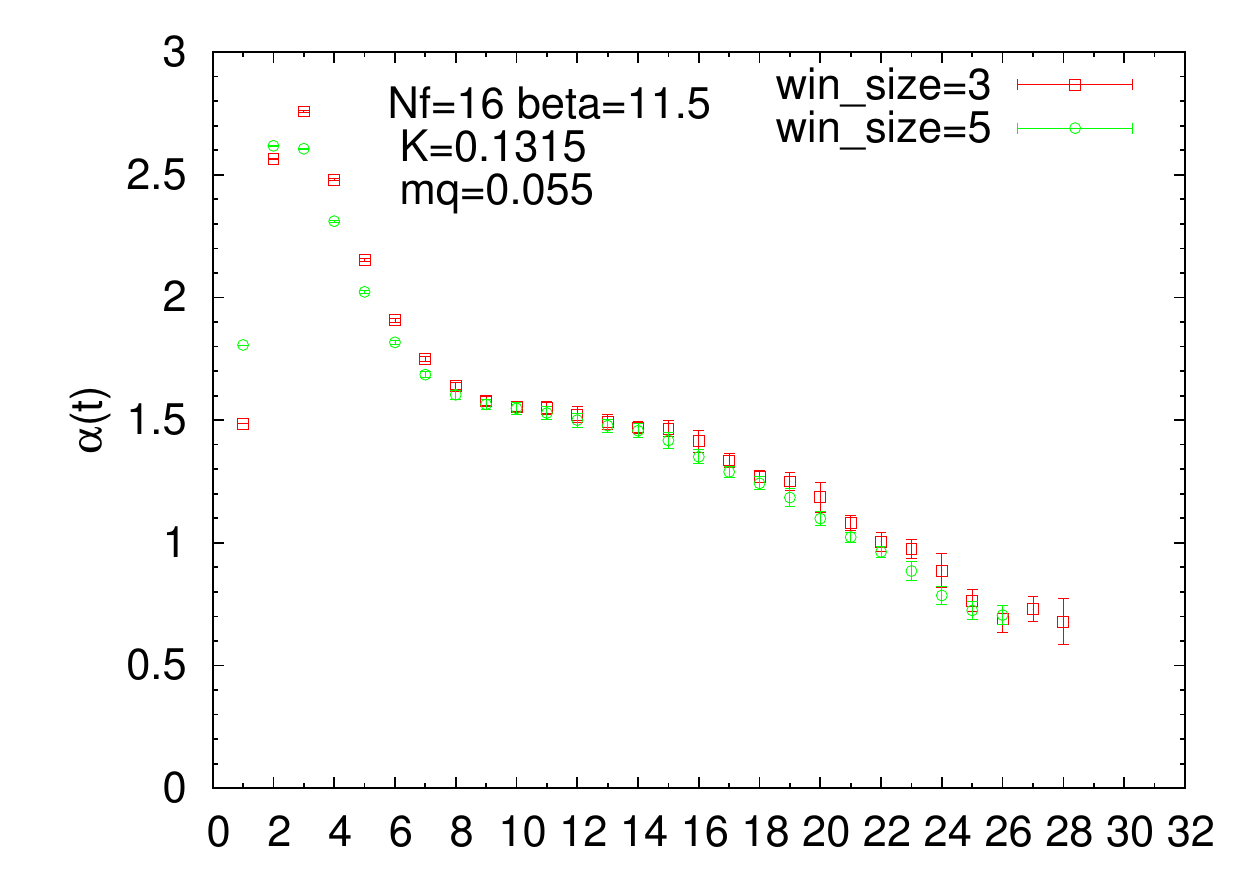}
\caption{ 
The correspondence of the local exponent $\alpha(t)$ for High Temperature QCD (left) and 
for Conformal QCD (right).}
\label{correspondence of exponent}
\end{figure*}

Let us show the results for the existence of the conformal region and the transition to a deconfining region is first order, in Figs.~\ref{nf16_effm} and \ref{nf16_comp} for the $N_f=16$ case.
Both of Fig.~\ref{nf16_effm} represent effective mass plots at the same $\beta$ and $K$; $\beta=11.5$ and $K=0.125$ ($m_q=0.24$).
The left one is the result taking a configuration at larger $K$ as an initial state, while the right one at smaller $K$.
We see that not only the values at $t \sim 30$ are quite different from each other, but the behavior at large $t$ are quite different. On the left there is no plateau up to $T=31$. We are able to fit the data with a power-law corrected Yukawa-type decaying form for the range $t=[15:31]$ with the exponent $\alpha=1.3(1)$.

Fig.~\ref{nf16_comp} represents the scattered plots of Polyakov loops in spacial directions overlaid.
The parameters are the same as Fig.~\ref{nf16_effm}. The left panel shows that the argument of the Polyakov loops are $2\pi /3$ with the absolute values $0.18$, while the right panel shows the arguments are $0$ and the absolute values are $0.05\sim 0.2$. These results clearly show the difference of the vacua of the two states.

\section{ Correspondence between Conformal QCD and High Temperature QCD }
Based on our theoretical analysis based on the RG flow and our
numerical simulations, we propose that there is a precise correspondence between
 Conformal QCD and High Temperature QCD and in the phase structure 
under the change of the parameters $N_f$ and $T/T_c$
with the same anomalous mass dimension.

We show the two sets of $\alpha(t)$
side by side in Figs.~\ref{correspondence of exponent},
the Conformal QCD data on the right panel
and the High Temperature QCD data on the left panel in order to compare them directly.
Here $\alpha(t)$ is a local exponent defined by
parametrizing the propagator $G(t)$ as
\begin{equation} 
G(t) = c\,  \frac {\exp(-m(t)\, t)}{t^{\, \alpha(t)}},
\label{local}
\end{equation}

We observe the correspondence on the $t$ dependence of $\alpha(t)$ between the two sets of data is excellent with the following each pair:
$T\sim 2 T_c$ and  $N_f=7$;
$T\sim 4 T_c$ and  $N_f=8$;
$T\sim 16 T_c$ and  $N_f=12$;
$T\sim 256T_c$ and  $N_f=16$.
Thus we plot schematically the correspondence between Conformal QCD and High Temperature QCD
as in Fig.(\ref{betaf2}). 
The correspondence is a powerful tool to investigate the properties of conformal theories.
The $T/T_c$ is a continuous variable, while the $N_f$ is a discrete variable.
Therefore we are able to use the information in High Temperature QCD to understand the properties of Conformal QCD.

Since High Temperature QCD covers  $0.0 \le \gamma^{*} \le 2.0$ and Conformal QCD takes
discrete values of $\gamma^*$  between $0.0$ and $2.0$ ($\gamma^*$ is the anomalous mass dimention),
the correspondence is realized between a continuous parameter $T/T_c$ and a discrete parameter $N_f$.
This is the precise origin of the correspondence between the two
 observed in the local-analysis of propagators.

The plateau at $15 \le t  \le 31$ in $\alpha(t)$ for $T \sim 2\, T_c$ disappears,
as the temperature increases to $T \sim 4\, T_c$.
Translating this fact into Conformal QCD is that the plateau in $\alpha(t)$ at $15 \le t \le 31$
observed  as  the IR behavior of $N_f=7$ disappears for $N_f=8$.

\subsection{$N_f=7$ and $T/T_c \simeq 2$}

We note that both in the $N_f=7$ case of Conformal QCD and 
at $T\sim 2 T_c$  in High Temperature QCD, 
we have 
a plateau in the $\alpha(t)$ at large $t$ ($15\,  \le t \le \, 31$).
In the both cases the IR behavior of the state is well described by a meson unparticle model~\cite{Cacciapaglia:2008ns}.

The value of $\alpha(t)$ at plateau($t=15\sim 31$) is $0.8(1)$ for $K=0.1452$ 
and $K=0.1459$ in the $N_f=7$ case.
Applying the formula $\alpha(t)=2 -\gamma^*$, we have
$\gamma^* = 1.2(1).$ 
Although this value  should be refined in the future by taking the continuum limit,
this value implies the anomalous mass dimension is of order unity.

\section*{Acknowledgments}
I would express gratitude to K.-I. Ishikawa, Yu Nakayama and T. Yoshie for a stimulating and
fruitful collaboration.
We would also like to thank 
S. Aoki, H. Fukaya, E. Itou, K. Kanaya, T Hatsuda, Y. Taniguchi, A. Ukawa and N. Yamada
for useful discussion.
The calculations were performed using HA-PACS computer at CCS, University of Tsukuba and SR16000
at KEK. I would like to thank members of CCS and KEK for their strong support for this work.


\end{document}